\documentclass[aps,twocolumn,superscriptaddress,longbibliography]{revtex4-2}
\usepackage[colorlinks=true,bookmarks=true,citecolor=magenta,urlcolor=magenta,linkcolor=magenta,breaklinks]{hyperref} 
\usepackage{breakurl}
\usepackage{xcolor, soul}
\usepackage{sidecap}
\usepackage{amssymb}
\usepackage{hhline}
\usepackage{multirow}
\sidecaptionvpos{figure}{t}
\usepackage{amsmath}
\usepackage{graphicx}
\usepackage{esint}
\usepackage{epstopdf}
\usepackage{rotating}
\usepackage{framed}
\graphicspath{{pict/}{./}}
\usepackage{bm}%
\usepackage{microtype,bm,bbm,graphicx,booktabs,times}

\newcounter{Fig}

\newcommand\mymapstol{\mathrel{\ooalign{$\leftarrow$\cr%
  \kern1.75ex\raise0.275ex\hbox{\scalebox{1}[0.4]{$\mid$}}\cr}}}

\newcommand\mymapstor{\mathrel{\ooalign{$\rightarrow$\cr%
  \kern-.15ex\raise.275ex\hbox{\scalebox{1}[0.4]{$\mid$}}\cr}}}
  
\usepackage{graphicx,tipa}

\usepackage{tikz}

\makeatletter
\newcommand*{\rom}[1]{\expandafter\@slowromancap\romannumeral #1@}
\makeatother
\begin{document}
\title{Polarization-Independent Zero Directional Scattering Without Geometric Symmetries}
\author{Chunchao Wen}
\author{Zhichun Qi}
\author{Jianfa Zhang}
\email{jfzhang85@nudt.edu.cn}

\author{Shiqiao Qin}
\email{sqqin8@nudt.edu.cn}
\author{Zhihong Zhu}
\author{Wei Liu}
\email{wei.liu.pku@gmail.com}
\affiliation{College for Advanced Interdisciplinary Studies, National University of Defense Technology, Changsha 410073, P. R. China.}
\affiliation{Nanhu Laser Laboratory and Hunan Provincial Key Laboratory of Novel Nano-Optoelectronic Information Materials and
Devices, National University of Defense Technology, Changsha 410073, P. R. China.}

\begin{abstract}
As the characteristic feature of generalized Kerker effect in Mie theory, directional scattering elimination has been playing a pivotal role in nanophotonics and many other  photonic disciplines, such as singular optics and topological photonics. Generally, zero directional scattering can be obtained only for a specific incident polarization, and to make it fully independent of arbitrary polarizations would require scatterers that exhibit geometric (\textit{e.g.} mirror) symmetries. Here we revisit the generalized Kerker effect and directional scattering elimination from the perspective of not the conventional electromagnetic multipoles, but rather quasi-normal modes supported by non-Hermitian systems. We reveal how to obtain zero directional scattering that is independent of arbitrary incident polarizations,  even for scattering structures that do not exhibit the required geometric symmetries. Such geometric symmetry-free and polarization-independent responses are made accessible through a synchronous exploitation of electromagnetic reciprocity and geometric phase. Our discovery can stimulate  fundamental explorations and practical applications in not only photonics, but also man other wave physics branches where scattering and geometric phase are pervasive.
\end{abstract}

\maketitle

\section{Introduction}
\label{section1}
Kerker effect and its generalized form featuring zero radiation along some open out-coupling channels have recently attracted broad interest and are widely employed for various applications in sensors, nanoantennnas, solar cells, radiative cooling devices, \textit{etc}~\cite{Kerker1983_JOSA,LIU_2018_Opt.Express_Generalized,KIVSHAR_2022_NanoLett._Rise,BABICHEVA_2024_Adv.Opt.Photon.AOP_Mieresonanta,YIN_2020_Science_Terrestrial}. Moreover, they have rapidly merged with the sweeping concepts of topology, singularity (in particular bound states in the continuum)  and non-hermiticity, revealing hidden interconnections and rendering exotic degrees freedom for light-matter interaction manipulations~\cite{LIU_ArXiv201204919Phys._Topological,MIRI_2019_Science_Exceptionala,KOSHELEV_2019_ScienceBulletin_Metaoptics,KANG_2023_NatRevPhys_Applications,WANG_2024_PhotonicsInsights_Optical}. For finite scattering bodies, Kerker effect (Kerker scattering) is manifest through zero directional scatterings along one or more directions, which is generically not robust against varying incident polarizations. To obtain arbitrary polarization-independent directional scattering eliminations, similar to other optical responses being irrelevant to polarizations,  geometric (mirror and/or larger than $2$-fold rotation) symmetries  of scatterers  seem to be inevitable~\cite{CHEN_2021_Phys.Rev.B_Arbitrary,YANG_LPR_Symmetry}.  Our central question is: \textit{is it possible to obtain Kerker scattering that is both independent of incident polarizations and free from geometric symmetries required for scattering bodies}?

The conventional interpretations of Kerker scattering are based on destructive interferences (along the corresponding zero scattering directions) among electromagnetic multipoles (spherical harmonics) of different orders and natures (electric and magnetic) ~\cite{Kerker1983_JOSA,LIU_2018_Opt.Express_Generalized}, which are of distinct even or odd phase parities~\cite{Liu2014_ultradirectional,LIU_Phys.Rev.Lett._generalized_2017}. Nevertheless, multipolar expansions are extrinsic: the  multipolar constituents are highly dependent on the origin of the reference frame chosen~\cite{Bohren1983_book,DOICU_light_2006,JACKSON_1998__Classical}, and for different incident polarizations the multipoles excited could be fully different  in terms of natures and orders. That is, the conventional language of spherical harmonics and electromagnetic multipoles sheds little light on our central question above, except for highly symmetric (such as spherical and cylindrical) structures of which multipoles themselves (with respect to the well-defined geometric centers of symmetric structures) happen to be eigen-modes supported.

Here we revisit Kerker scattering from the perspective of excitations of quasi-normal modes (QNMs) and their mutual interferences~\cite{LALANNE__LaserPhotonicsRev._Light,KRISTENSEN_2020_Adv.Opt.Photon.AOP_Modeling}. In sharp contrast to multipolar expansions, QNM expansions are intrinsic (independent of the reference frame) and for varying incident polarizations, the QNMs involved are generally fixed (eigen-modes supported by the scatterer in the spectral regime of interest), except for different excitation efficiencies and (geometric) phases~\cite{LALANNE__LaserPhotonicsRev._Light,KRISTENSEN_2020_Adv.Opt.Photon.AOP_Modeling,CHEN_Phys.Rev.Lett._Extremize,WANG_2024_Phys.Rev.Lett._Geometric}. We reveal two scenarios when geometric symmetry-free arbitrary polarization-independent zero directional scattering (along $\hat{\mathbf{r}}_\mathrm{o}$) can be obtained: (i) The radiations of all QNMs excited are zero along $\hat{\mathbf{r}}_\mathrm{o}$; (ii) QNM radiations along $\hat{\mathbf{r}}_\mathrm{o}$ are not zero, but interfere destructively and fully cancel out, irrespective of the incident polarizations. The second scenario is accessible when QNM radiations opposite to the incident direction  are of the same polarization, which secures that the relative excitation efficiencies and geometric phases among QNMs are polarization-independent. The principles we reveal to achieve polarization-independent zero directional scattering are robustly protected by electromagnetic reciprocity, which will further accelerate and broaden the applications of Kerker effects across different disciplines of photonics.

\section{Zero directional scattering from the perspective of QNMs}
\label{section2}

\begin{figure}[tp]
\centerline{\includegraphics[width=8.5cm]{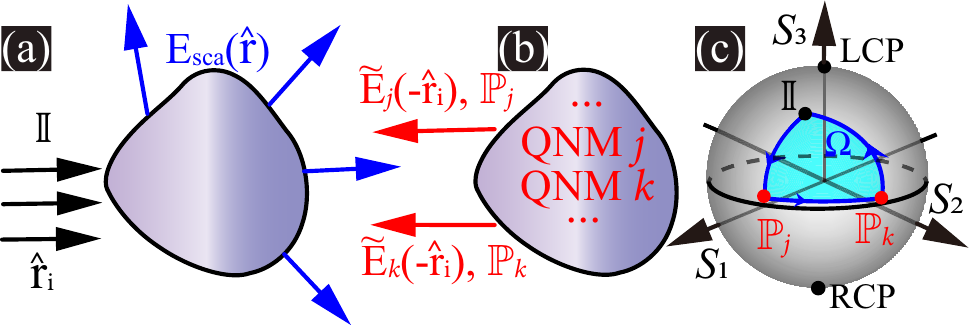}} \caption {\small (a) For a plane wave propagating along the direction $\hat{\mathbf{r}}_\mathrm{i}$ with polarization $\mathbb{I}$ [see its position on the Poincar\'{e} sphere in (c)], the scattered wave along  $\mathbf{\hat{r}}$ is $\tilde{\mathbf{E}}_{\rm{sca}}(\hat{\mathbf{r}})$. (b) The scatterer supports a discrete set of QNMs (indexed by the integer $m$) and the far field radiation of the $m^{\mathrm{th}}$ QNM opposite to the incident direction is $\tilde{\mathbf{E}}_m(-\hat{\mathbf{r}}_\mathrm{i})$, with the corresponding polarization $\mathbb{P}_m$. The relative geometric phase between QNMs $j$ and $k$ is represented geometrically on the Poincar\'{e} sphere trhough a geodesic triangle $\triangle\mathbb{I}\mathbb{P}_j\mathbb{P}_k$ enclosing a solid angle $\mathbf{\Omega}$: $\varphi_{j,k}=\frac{1}{2}\mathbf{\Omega}$. Here LCP and RCP denote respectively left-handend and right-handed circularly polarized light.}\label{fig1}
\end{figure}

For incident plane waves (propagating along $\hat{\mathbf{r}}_\mathrm{i}$) [Fig.~\ref{fig1} (a)], the scattering process can be divided into two steps from the perspective of QNMs: (i) The incident wave excites a discrete set of QNMs supported by the scatterer (indexed by positive integer $m$) with complex eigenfrequencies $\tilde{\omega}_m$ and eigenfields $\tilde{\mathbf{E}}_m(\mathbf{r})$~\cite{LALANNE__LaserPhotonicsRev._Light,KRISTENSEN_2020_Adv.Opt.Photon.AOP_Modeling}; (ii) The QNMs excited then radiate to all directions, and the summed radiations together with the incident field account for various scattering properties such as cross sections of scattering, extinction and absorption~\cite{Bohren1983_book,DOICU_light_2006,JACKSON_1998__Classical}. The far-field radiations of QNMs [denoted as $\tilde{\mathbf{E}}_m(\mathbf{\hat{r}})$ and $\mathbf{\hat{r}}$ is the unit direction vector] are transverse and the corresponding radiation polarization can be described by either the Jones vector or Stokes vector $\mathbf{S}$ (with three components $S_{1,2,3}$) on the Poincar\'{e} sphere at point $\mathbb{P}_m(\mathbf{\hat{r}})$ [Figs.~\ref{fig1} (b) and \ref{fig1} (c)]~\cite{YARIV_2006__Photonics}. In the far field, the scattered wave $\mathbf{E}_{\rm{sca}}(\hat{\mathbf{r}})$ can be expanded into QNM radiations as~\cite{LALANNE__LaserPhotonicsRev._Light,KRISTENSEN_2020_Adv.Opt.Photon.AOP_Modeling}:
\begin{equation}
\label{expansion}
\mathbf{E}_{\rm{sca}}(\hat{\mathbf{r}})=\sum_m  \alpha_{m} {\tilde{\mathbf{E}}}_{m}(\hat{\mathbf{{r}}}),  
\end{equation}
where $\alpha_{m}$ is the  complex excitation (coupling) coefficient for the $m^ {\mathrm{th}}$ QNM under the incident plane wave, and  the incident polarization is represented as point  $\mathbb{I}$ on the Poincar\'{e} sphere  [Fig.~\ref{fig1} (c)].

For reciprocal scatterers,  the excitation coefficients can be calculated in the far field, and thus simplified as (with unit incident electric field vector)~\cite{CHEN_Phys.Rev.Lett._Extremize,WANG_2024_Phys.Rev.Lett._Geometric}:

\begin{equation}
\label{coefficient-geometric}
\alpha_{m}=\cos\left(\frac{1}{2}\overset{\smash{\raisebox{0.01ex}{\tikz\draw (0,0) arc[start angle=120,end angle=60,radius=0.3cm];}}}{\mathbb{I}\mathbb{P}}_m(-\hat{\mathbf{r}}_\mathrm{i})\right)\exp({i\varphi_m}).
\end{equation}
As already specified,  $\mathbb{I}$ and $\mathbb{P}_m(-\hat{\mathbf{r}}_\mathrm{i})$ are points on the unit Poincaré sphere, denoting respectively the incident polarization and the radiation polarization (for the $m^ {\mathrm{th}}$ QNM) opposite to the incident direction [see Figs.~\ref{fig1}(b) and \ref{fig1}(c); in the following discussions we drop the term $(-\hat{\mathbf{r}}_\mathrm{i})$ for simplicity];  $\overset{\smash{\raisebox{0.01ex}{\tikz\draw (0,0) arc[start angle=120,end angle=60,radius=0.3cm];}}}{\mathbb{I}\mathbb{P}}_m$ is the geodesic distance between these two points  [Fig.~\ref{fig1}(c)]; $\exp({i\varphi_m})$ is the overall phase factor~\cite{WANG_2024_Phys.Rev.Lett._Geometric}. The above Eq.~(\ref{coefficient-geometric}) tells that, except for the incident wave, the excitation efficiency for each QNM is solely related to its radiation opposite to the incident direction, with the term $\cos(\frac{1}{2}\overset{\smash{\raisebox{0.01ex}{\tikz\draw (0,0) arc[start angle=120,end angle=60,radius=0.3cm];}}}{\mathbb{I}\mathbb{P}}_m)$ characterizing the polarization difference between the incident wave and the QNM radiation: $\cos(\frac{1}{2}\overset{\smash{\raisebox{0.01ex}{\tikz\draw (0,0) arc[start angle=120,end angle=60,radius=0.3cm];}}}{\mathbb{I}\mathbb{P}}_m)=1$  ( $\mathbb{I}$ and $\mathbb{P}_m$ are overlapped) and $\cos(\frac{1}{2}\overset{\smash{\raisebox{0.01ex}{\tikz\draw (0,0) arc[start angle=120,end angle=60,radius=0.3cm];}}}{\mathbb{I}\mathbb{P}}_m)=0$ (  $\mathbb{I}$ and $\mathbb{P}_m$ are diametrically opposite antipodal points $\overset{\smash{\raisebox{0.01ex}{\tikz\draw (0,0) arc[start angle=120,end angle=60,radius=0.3cm];}}}{\mathbb{I}\mathbb{P}}_m=\pi$) represent fully matched and orthogonal polarizations, respectively. 

Along an arbitrary scattering direction, the radiations of all QNMs interfere and thus the relative phase between them are equally important. For any pair of QNMs (indexed by $j$ and $k$), the relative phase can be simplified as a pure geometric phase~\cite{WANG_2024_Phys.Rev.Lett._Geometric}:
\begin{equation}
\label{geometric-phase}
\varphi_{j,k}=\varphi_j-\varphi_k=\frac{1}{2}\mathbf{\Omega}(\triangle\mathbb{I}\mathbb{P}_j\mathbb{P}_k),
\end{equation}
where $\mathbf{\Omega}(\triangle\mathbb{I}\mathbb{P}_j\mathbb{P}_k)$ denotes the signed  solid angle enclosed by the geodesic triangle $\triangle\mathbb{I}\mathbb{P}_j\mathbb{P}_k$ [positive (negative) for a counter-clockwise (clockwise) circuit transversing $\mathbb{I}$, $\mathbb{P}_j$, $\mathbb{P}_k$ and back to $\mathbb{I}$ consecutively; see Fig.~\ref{fig1} (c)]. The above Eqs.~(\ref{expansion})-(\ref{geometric-phase}) clarify immediately why optical scattering properties are generally dependent on incident polarizations: with varying polarizations, the excitation efficiency for each QNM and the relative phase between any pair of QNMs would generally change, which would induce variations of QNM interferences and thus the overall scattering features. 

To obtain zero directional scattering along  $\hat{\mathbf{r}}_\mathrm{o}$ that is independent of the incident polarizations, it requires that for an arbitrary point $\mathbb{I}$ on the Poincar\'{e} sphere:
\begin{equation}
\label{zero-directional}
\mathbf{E}_{\rm{sca}}(\hat{\mathbf{r}}_\mathrm{o})=\sum_m  \cos\left(\frac{1}{2}\overset{\smash{\raisebox{0.01ex}{\tikz\draw (0,0) arc[start angle=120,end angle=60,radius=0.3cm];}}}{\mathbb{I}\mathbb{P}}_m\right)\exp({i\varphi_m}) {\tilde{\mathbf{E}}}_{m}(\hat{\mathbf{r}}_\mathrm{o})=0.  
\end{equation}
One sufficient condition is that  the radiations of all QNMs excited are zero along $\hat{\mathbf{r}}_\mathrm{o}$: ${\tilde{\mathbf{E}}}_{m}(\hat{\mathbf{r}}_\mathrm{o})=0$. Another scenario is that  all $\mathbb{P}_m$ overlap at  $\mathbb{P}_\mathrm{o}$ [the radiations of all QNMs excited opposite to the incident direction are of the same polarization; $\varphi_{j,k}=0$ in Eq.~(\ref{geometric-phase}) and thus the phase factor $\exp({i\varphi_m})$ converges to  $\exp({i\varphi_\mathrm{o}})$], which simplifies Eq.~(\ref{zero-directional}) to:
\begin{equation}
\label{zero-directional-sim}
\mathbf{E}_{\rm{sca}}(\hat{\mathbf{r}}_\mathrm{o})=\cos\left(\frac{1}{2}\overset{\smash{\raisebox{0.01ex}{\tikz\draw (0,0) arc[start angle=120,end angle=60,radius=0.3cm];}}}{\mathbb{I}\mathbb{P}}_\mathrm{o}\right) \exp({i\varphi_\mathrm{o}}) \sum_m  {\tilde{\mathbf{E}}}_{m}(\hat{\mathbf{r}}_\mathrm{o})=0.  
\end{equation}
It means that as long as  $\sum_m  {\tilde{\mathbf{E}}}_{m}(\hat{\mathbf{r}}_\mathrm{o})=0$, the direction scattering  along $\hat{\mathbf{r}}_\mathrm{o}$ would be invariantly zero irrespective of incident polarizations (positions of $\mathbb{I}$ on the Poincar\'{e} sphere).  In the elementary single-QNM regime (when only an individual QNM is supported in the spectral regime of interest), the two scenarios of ${\tilde{\mathbf{E}}}_{m}(\hat{\mathbf{r}}_\mathrm{o})=0$ and $\sum_m  {\tilde{\mathbf{E}}}_{m}(\hat{\mathbf{r}}_\mathrm{o})=0$ are essentially the same. The simplest example is  a metal cylinder supporting only an electric dipolar mode: no matter what the incident directions and polarizations are, the scattering is always zero along the two directions parallel to the cylinder axis.

In the following two sections, we will verify numerically those principles and exemplify the two corresponding scenarios [${\tilde{\mathbf{E}}}_{m}(\hat{\mathbf{r}}_\mathrm{o})=0$ for all $m$ in Section~\ref{section3}; $\sum_m  {\tilde{\mathbf{E}}}_{m}(\hat{\mathbf{r}}_\mathrm{o})=0$ with overlapped $\mathbb{P}_m$ in Section~\ref{section4}] with specific scattering configurations. Considering that all principles revealed above are irrelevant to the number of the QNMs co-exited, without loss of generality we focus on simultaneous excitations of two QNMs only.

\section{Polarization-independent zero directional scattering induced by overlapped zero QNM radiations}
\label{section3}

\begin{figure}[btp]
	\centerline{\includegraphics[width=0.4\textwidth]{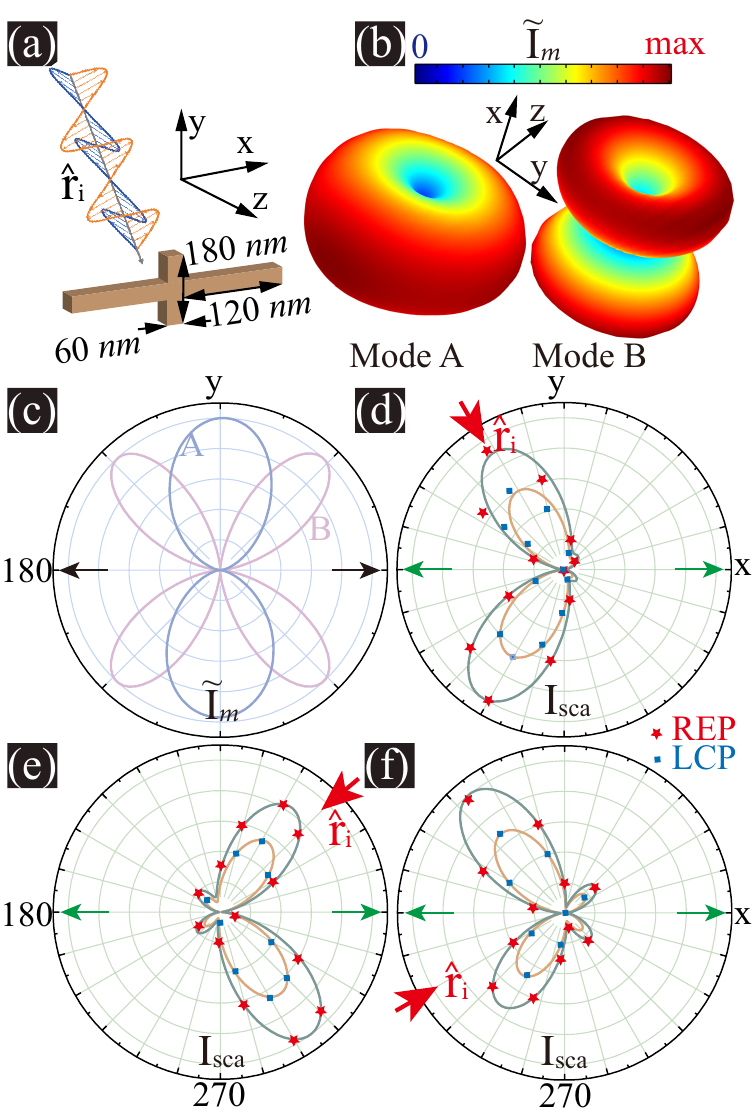}} \caption {\small (a)~A gold scatterer with all geometric parameters specified. (b)~3D and (c)~2D (on the \textbf{x}-\textbf{y} plane) angular radiation patterns of two QNMs supported. In (c) the directions of overlapped zero radiation directions have been marked by black arrows. (d)-(f) Angular scattering patterns for two incident polarizations (LCP with $S_3$=1 and REP with $S_3$=-$\sqrt{2}/2$) for three randomly chosen incident directions. In (d)-(f) these incident directions (red arrows) and zero directional scattering (green arrows) have also been marked.}\label{fig2}
\end{figure} 

We start with a simple symmetric (exhibiting mirror and inversion symmetry) plasmonic gold structure, with all geometric parameters specified in Fig.~\ref{fig2}(a). For the  relative permittivity of gold, we adopt the Drude model to fit the experimental data~\cite{Johnson1972_PRB}:  $\epsilon_r(\omega)=1-\omega_{p}^{2}/(\omega^2+i\Gamma\omega)$ with plasma frequency $\omega_p=1.37\times10^{16}~\mathrm{rad}/s$  and collision frequency $\Gamma=8.17\times10^{13}~\mathrm{rad}/s$. Numerical results throughout this work are obtained by full-wave electromagnetic simulations in COMSOL Multiphysics, for both calculations with (scattering properties) and without (QNM properties) incident plane waves. Over the spectral regime of interest, two QNMs A and B are supported  with complex eigenfrequencies: $\widetilde{\omega}_{\rm{A}}=(1.505\times 10^{15}+5.272\times10^{14}\textit{i})~\mathrm{rad}/s$ and $\widetilde{\omega}_{\rm{B}}=(1.447\times 10^{15}+5.776\times10^{13}\textit{i}~\mathrm{rad}/s)$. The corresponding far-field 3D (three-dimensional) and 2D (two-dimensional; on the \textbf{x}-\textbf{y} plane) QNM radiation patterns [$\tilde{\mathbf{I}}_m(\mathbf{r})\propto|\tilde{\mathbf{E}}_m(\mathbf{r})|^2$]  are shown  in Figs.~\ref{fig2}(b) and~\ref{fig2}(c), respectively.  Notably, zero radiation directions of both QNMs overlap at two opposite directions, as indicated by black arrows in Fig.~\ref{fig2}(c). Then according to Eq.~(\ref{zero-directional}), irrespective of incident directions and polarizations, scattering along those indicated directions would be invariantly zero. This is further verified for three incident directions (indicated by red arrows) with incident wavelength ${\lambda}_\mathrm{i}=1250$~nm (angular frequency ${\omega}_{\rm{i}}=1.5069\times 10^{15}~\mathrm{rad}/s$), as seen in  Figs.~\ref{fig2}(d)-\ref{fig2}(f) showing the corresponding angular scattering patterns[$\mathbf{I}_{\rm{sca}}(\hat{\mathbf{r}})\propto|\mathbf{E}_{\rm{sca}}(\hat{\mathbf{r}})|^2]$. For each incident direction, two sets of results are shown for LCP ($S_3=1$) and REP (right-handed elliptically polarized; $S_3=-\sqrt{2}/2$) incidences, respectively. As it is clearly shown, for all scenarios, the directional scattering along the directions indicated in  Fig.~\ref{fig2}(c) is eliminated.  

\begin{figure}[btp]
	\centerline{\includegraphics[width=0.4\textwidth]{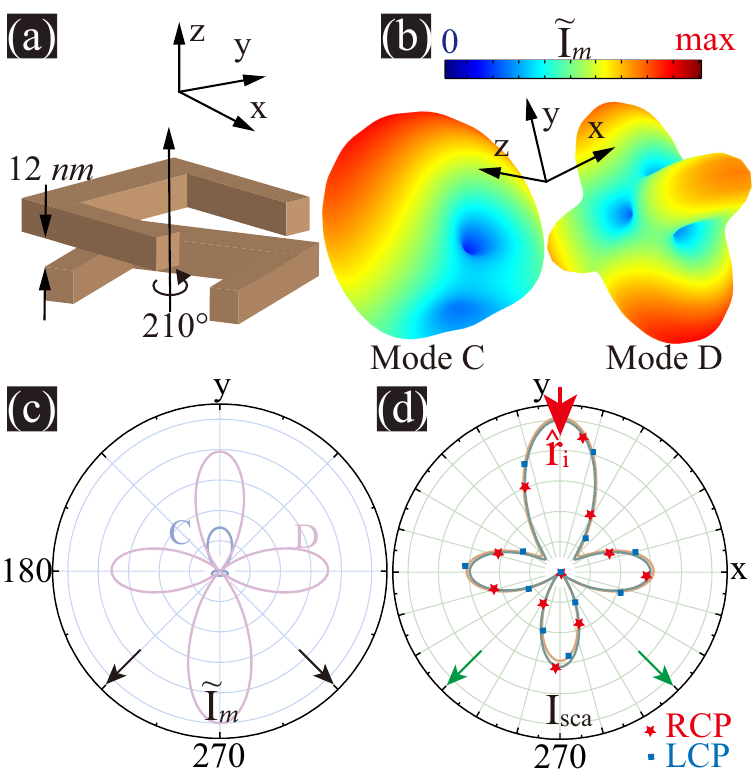}} \caption {\small(a)~A pair of coupled identical parallel SRRs (without mirror or rotation symmetries) with a displacement $12$~nm and a twist angle $210^\circ$ along the \textbf{z}-axis. The parameters of the indiviudal SRR are specified in Fig.~\ref{fig4}(a). (b)~3D and (c)~2D  angular radiation patterns of two QNMs. The directions of overlapped zero radiations (black arrows) have been marked. (d)~Angular scattering patterns for two circular incident polarizations. The incident direction (red arrow) and directions of  zero scattering (green arrows) have all been marked.}\label{fig3}
\end{figure} 

We then turn to another structure [coupled split ring resonators (SRRs); refer to Fig.~\ref{fig3}(a) for the geometric parameters of individual resonators] exhibiting no mirror, inversion or rotation symmetry, as shown in Fig.~\ref{fig3}(a). The two SRR are parallel-displaced by a distance of $12$~nm and twisted by angle of $210^{\circ}$ with respect to each other along \textbf{z}-axis. Two QNMs C and D are supported with $\widetilde{\omega}_{\rm{C}}=(2.092\times 10^{15}+4.449\times10^{13}\textit{i})$~rad/s and $\widetilde{\omega}_{\rm{D}}=(2.473\times 10^{15}+2.817\times10^{13}\textit{i})$~rad/s. Their corresponding 3D and 2D radiation patterns are shown in Figs.~\ref{fig3}(b) and~\ref{fig3}(c) respectively.  Similar to the symmetric structure shown in Fig.~\ref{fig2}, there are two directions along which both QNM radiations are zero [indicated in Fig.~\ref{fig3}(c)]. Zero directional scattering along those directions are verified in Fig.~\ref{fig3}(d), for both RCP and LCP incidences with incident wavelength ${\lambda}_\mathrm{i}=790$~nm (angular frequency ${\omega}_{\rm{i}}=2.3844\times 10^{15}~\mathrm{rad}/s$).

\section{Polarization-independent zero directional scattering induced by destructive interferences of QNM radiations}
\label{section4}

\begin{figure}[btp]
	\centerline{\includegraphics[width=0.4\textwidth]{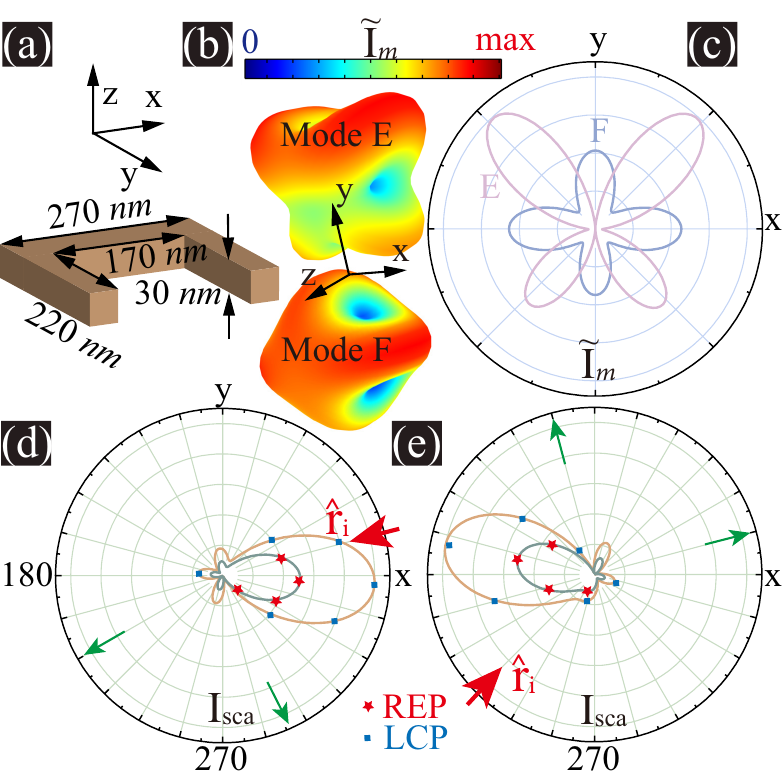}} \caption {\small (a)~A symmetric gold SRR with all geometric parameters specified. (b)~3D and (c)~2D angular radiation patterns of two QNMs E and F supported by this SRR.  (d)-(e) Angular scattering patterns for two incident polarizations (LCP with $S_3$=1 and REP with $S_3$=-$\sqrt{2}/2$) for two incident directions (red arrows), with two  zero directional scattering directions (green arrows) for each case.}\label{fig4}
\end{figure} 

As a next step, we proceed to the second scenario of zero directional scattering, not induced by overlapped zero radiation of all QNMs, but by non-zero QNM radiations that all cancel out due to destructive interferences.  Similar to  the last section,  we start with a symmetric individual SRR with all geometric parameters specified in Fig.~\ref{fig4}(a). This SRR supports a pair of quadrupole-like QNMs E and F with eigenfrequencies $\widetilde{\omega}_{\rm{E}}=(3.061\times 10^{15}+2.545\times10^{12}\textit{i}~\mathrm{rad}/s)$ and $\widetilde{\omega}_{\rm{F}}=(2.25\times 10^{15}+3.786\times10^{13}\textit{i}~\mathrm{rad}/s)$. Their corresponding 3D and 2D  far-field radiation patterns are shown in Figs.~\ref{fig4}(b) and~\ref{fig4}(c) respectively. For these two QNMs: (i) On the \textbf{x}-\textbf{y} plane, both QNM radiations are of an identical linear polarization ($S_3 = 0$) orientated along \textbf{z}-axis ($\mathbb{P}_\mathrm{E}$ and $\mathbb{P}_\mathrm{F}$ overlap on the equator of the Poincar\'{e} sphere); (ii) QNM radiation polarizations for all other directions out of  the \textbf{x}-\textbf{y} plane. Then according to  Eq.~(\ref{zero-directional-sim}), for an arbitrary incident direction on the \textbf{x}-\textbf{y} plane, the zero directional scattering obtained would be polarization-independent. 
The angular scattering for two different incident directions on the \textbf{x}-\textbf{y} plane are showcased in Figs.~\ref{fig4}(e) and~\ref{fig4}(f) [two incident polarizations for each incident direction with incident wavelength ${\lambda}_\mathrm{i}=750$~nm (angular frequency ${\omega}_{\rm{i}}=2.5115\times 10^{15}~\mathrm{rad}/s$)], and all zero-scattering directions are indicated by green arrows. Compared to the case studies in Section~\ref{section3}, here along those directions the QNM radiations are not zero and thus zero scattering originates from destructive interferences between them.
\begin{figure}[b]
	\centerline{\includegraphics[width=0.38\textwidth]{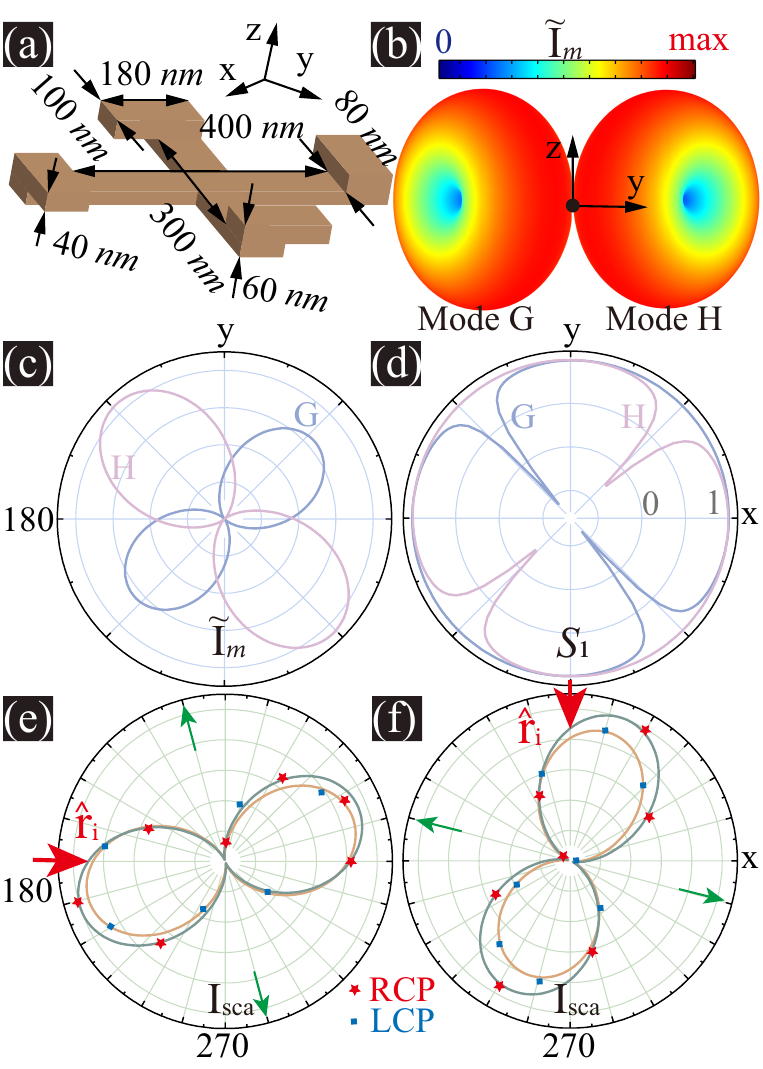}} \caption {\small (a)~A gold particle exhibiting no geometric (mirror or larger than $2$-fold rotation) symmetries with all geometric parameters specified. (b)~3D and (c)~2D angular radiation patterns, and (d) radiation polarization distributions (in terms of $\mathbf{S}_1$) for the two supported QNMs G and H. In (d) four directions of identical QNM radiation polarizations are marked by red dots.  Angular scattering patterns of this scatterer for  incident waves (LCP and RCP) along two such directions are shown in (e) and (f), and for each case two directions of polarization-independent zero-scattering directions are marked by green arrows.}\label{fig5}
\end{figure} 

We have further investigated another asymmetric gold scatterer (without mirror, inversion or more than $2$-fold rotation symmetry) schematically shown in Fig.~\ref{fig5}(a), which supports a pair of dipole-like QNMs G and H.  Complex eigenfrequencies of these two QNMs are $\widetilde{\omega}_{\rm{G}}=(7.453\times 10^{14}+1.583\times10^{13}\textit{i})~\mathrm{rad}/s$ and $\widetilde{\omega}_{\rm{H}}=(8.498\times 10^{14}+1.210\times10^{13}\textit{i})~\mathrm{rad}/s$. Their corresponding 3D and 2D  far-field radiation patterns are shown in Figs.~\ref{fig5}(b) and~\ref{fig5}(c), respectively.  Distinct from the symmetric structure in Fig.~\ref{fig4} for which there are an infinite number of directions where QNM radiation polarizations are identical, on the \textbf{x}-\textbf{y} plane there are only two pairs of directions with the same QNM polarizations: the polarization distributions of two QNMs on the \textbf{x}-\textbf{y} plane (in terms of Stokes parameter $\mathbf{S}_1$; polarizations are linear with $\mathbf{S}_3=0$) are further shown in Fig.~\ref{fig5}(d) and the directions of overlapped $\mathbb{P}_\mathrm{G}$ and $\mathbb{P}_\mathrm{H}$ are pinpointed (red dots). Along all four directions the QNM radiations are obviously not zero.  The angular scattering patterns for waves incident opposite to such directions are showcased in Figs.~\ref{fig5}(e) and~\ref{fig5}(f) [two incident polarizations for each incident direction with incident wavelength ${\lambda}_\mathrm{i}=2.540~\mu$m (angular frequency ${\omega}_{\rm{i}}=7.416\times 10^{14}~\mathrm{rad}/s$)], and polarization-independent zero-scattering directions are indicated by green arrows. 
\begin{figure}[btp]
	\centerline{\includegraphics[width=0.45\textwidth]{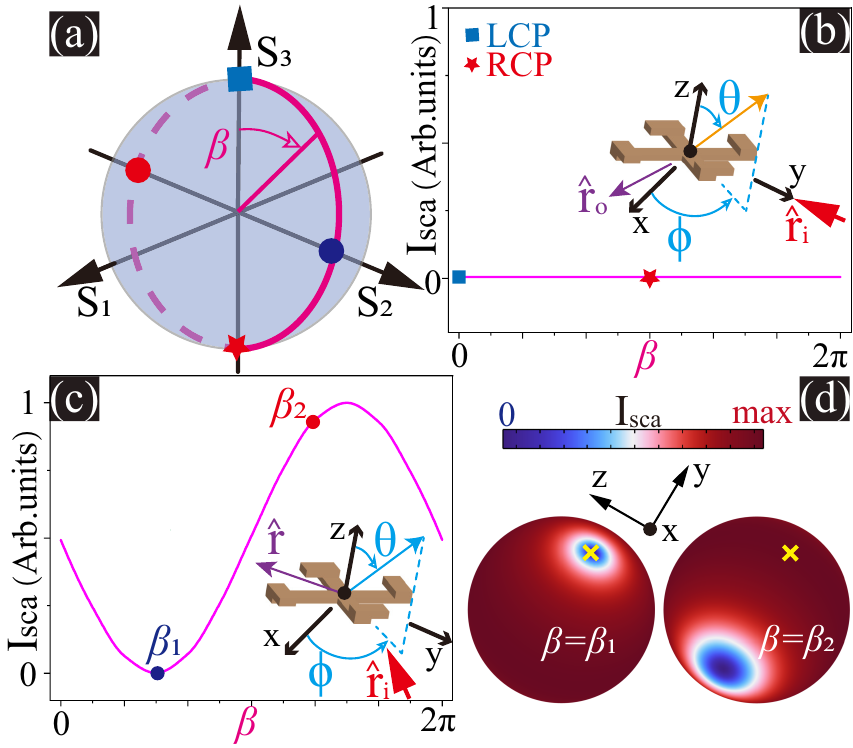}} \caption {\small (a)~Incident polarizations transverse a great circle of the Poincar\'{e} sphere that is parametrized by $\beta$. 
		The dependence of directoinal scattering intensity on $\beta$ for the same scatterer shown in Fig.~\ref{fig5}: (b) The incident and scattering directions are  that same as those marked  in Fig.~\ref{fig5}(f); (c) The incident direction is $\theta=150^\circ$, $\phi=90^\circ$ opposite to which $\mathbb{P}_\mathrm{G}$ and $\mathbb{P}_\mathrm{H}$ are not overlapped,  and the marked scattering direction  is $\theta=116.2^\circ$, $\phi=22.4^\circ$. The scattering reaches zero at $\beta=\beta_1=\pi/2$, and generally are not zero at other 
positions (\textit{e.g.} for $\beta=\beta_2=4\pi/3$). (d) The 3D scattering patterns for two incident polarizations indicated in (c), and the directional scattering along the marked direction is obviously zero and not zero at $\beta_1$ and $\beta_2$, respectively.}\label{fig6}
\end{figure} 
Finally, to further illustrate the arbitrary-polarization independence of the zero directional scattering obtained, we have shown the evolutions of directional scattering intensity for incident polarizations transversing a great circle of the Poincar\'{e} sphere parameterized by $\beta$ [Fig.~\ref{fig6}(a)]: $\beta=0$ for LCP and $\beta=\pi$ for RCP.  For the results shown in Figs.~\ref{fig6}(b) and (c), the scattering particle and incident wavelength is the same as that in Fig.~\ref{fig5}: in Fig.~\ref{fig6}(b) the incident and scattering directions are the same as that marked in Fig.~\ref{fig5}(f). As is clearly shown, the directional scattering is invariantly zero for any incident polarizations; for comparison in  Fig.~\ref{fig6}(c), we have chosen another incident direction ($\theta=150^\circ$, $\phi=90^\circ$), opposite to which  the QNM radiation polarizations are not identical ($\mathbb{P}_\mathrm{G}$ and $\mathbb{P}_\mathrm{H}$ are not overlapped). For such a incident direction and the selected scattering direction  ($\theta=116.2^\circ$, $\phi=22.4^\circ$), the dependence of directional scattering intensity on $\beta$ shown in Fig.~\ref{fig6}(c) verifies that though scattering can be zero for one polarization ($\beta=\pi/2$), it is generally not zero for all other polarizations (\textit{e.g.} for $\beta=\beta_2=4\pi/3$). In other words, the zero directional scattering obtained is polarization-dependent. This is due to the fact  the  $\mathbb{P}_\mathrm{G}$ and $\mathbb{P}_\mathrm{H}$ are not overlapped, and thus the relative phases and amplitudes between the QNMs excited are varying for different incident polarizations [see Fig.~\ref{fig1}(c)], which ensures the full destructive interference among QNM radiations for only one specific incident polarization [see Eq.~(\ref{zero-directional})].  Such polarization dependence is further showcased through 3D scattering patterns in Fig.~\ref{fig6}(d): along the selected scattering direction, the scattering is zero for $\beta=\pi/2$ and nonzero for $\beta=4\pi/3$.\\

\section{Conclusion}

To conclude,  we have merged two sweeping concepts, Kerker scattering in Mie theory and quasi-normal modes in non-Hermitian photonics, based on which we obtain polarization-independent zero directional scattering, even for scatterers that do not exhibit geometric (mirror or more than two-fold rotation) symmetries. Such geometry symmetry-free polarization-independent optical responses have been made accessible through a synchronous exploitation of electromagnetic reciprocity and geometric phase.  Zero directional scattering is a bridging phenomena among the vibrant topics including geometric phase, singular optics, topological photonics. We thus expect our discovery, which has essentially revealed a deeper hidden connection, can stimulate both fundamental explorations and practical applications in many related disciplines, in not only photonics but also general wave physics where scattering and geometric phase are pervasive,  opening extra dimensions of freedom to exploit for extreme wave-matter interaction manipulations.  \\

\section*{acknowledgement}
This research was funded by the National Natural Science Foundation of
China (12274462, 11674396, and 11874426) and several other projects of Hunan Province (2024JJ2056, 2018JJ1033 and 2017RS3039). 




\begin{thebibliography}{10}
\newcommand{\enquote}[1]{``#1''}

\bibitem{Kerker1983_JOSA}
M.~Kerker, D.~S. Wang, and C.~L. Giles, \enquote{Electromagnetic scattering by
  magnetic spheres,} J. Opt. Soc. Am. \textbf{73}, 765 (1983).

\bibitem{LIU_2018_Opt.Express_Generalized}
W.~Liu and Y.~S. Kivshar, \enquote{Generalized {{Kerker}} effects in
  nanophotonics and meta-optics {{[Invited]}},} Opt. Express \textbf{26},
  13085--13105 (2018).

\bibitem{KIVSHAR_2022_NanoLett._Rise}
Y.~Kivshar, \enquote{The {{Rise}} of {{Mie-tronics}},} Nano Lett. \textbf{22},
  3513--3515 (2022).

\bibitem{BABICHEVA_2024_Adv.Opt.Photon.AOP_Mieresonanta}
V.~E. Babicheva and A.~B. Evlyukhin, \enquote{Mie-resonant metaphotonics,} Adv.
  Opt. Photon., AOP \textbf{16}, 539--658 (2024).

\bibitem{YIN_2020_Science_Terrestrial}
X.~Yin, R.~Yang, G.~Tan, and S.~Fan, \enquote{Terrestrial radiative cooling:
  {{Using}} the cold universe as a renewable and sustainable energy source,}
  Science \textbf{370} (2020).

\bibitem{LIU_ArXiv201204919Phys._Topological}
W.~Liu, W.~Liu, L.~Shi, and Y.~Kivshar, \enquote{Topological polarization
  singularities in metaphotonics,} Nanophotonics \textbf{10}, 1469--1486
  (2021).

\bibitem{MIRI_2019_Science_Exceptionala}
M.-A. Miri and A.~Al{\`u}, \enquote{Exceptional points in optics and
  photonics,} Science \textbf{363}, eaar7709 (2019).

\bibitem{KOSHELEV_2019_ScienceBulletin_Metaoptics}
K.~Koshelev, A.~Bogdanov, and Y.~Kivshar, \enquote{Meta-optics and bound states
  in the continuum,} Science Bulletin \textbf{64}, 836--842 (2019).

\bibitem{KANG_2023_NatRevPhys_Applications}
M.~Kang, T.~Liu, C.~T. Chan, and M.~Xiao, \enquote{Applications of bound states
  in the continuum in photonics,} Nat Rev Phys \textbf{5}, 659--678 (2023).

\bibitem{WANG_2024_PhotonicsInsights_Optical}
J.~Wang, P.~Li, X.~Zhao, Z.~Qian, X.~Wang, F.~Wang, X.~Zhou, D.~Han, C.~Peng,
  L.~Shi, and J.~Zi, \enquote{Optical bound states in the continuum in periodic
  structures: Mechanisms, effects, and applications,} Photonics Insights
  \textbf{3}, R01 (2024).

\bibitem{CHEN_2021_Phys.Rev.B_Arbitrary}
W.~Chen, Q.~Yang, Y.~Chen, and W.~Liu, \enquote{Arbitrary
  polarization-independent backscattering or reflection by rotationally
  symmetric reciprocal structures,} Phys. Rev. B \textbf{103}, 045422 (2021).

\bibitem{YANG_LPR_Symmetry}
Q.~Yang, W.~Chen, Y.~Chen, and W.~Liu, \enquote{Symmetry {{Protected Invariant
  Scattering Properties}} for {{Incident Plane Waves}} of {{Arbitrary
  Polarizations}},} Laser \& Photonics Reviews \textbf{15}, 2000496 (2021).

\bibitem{Liu2014_ultradirectional}
W.~Liu, J.~Zhang, B.~Lei, H.~Ma, W.~Xie, and H.~Hu, \enquote{Ultra-directional
  forward scattering by individual core-shell nanoparticles,} Opt. Express
  \textbf{22}, 16178 (2014).

\bibitem{LIU_Phys.Rev.Lett._generalized_2017}
W.~Liu, \enquote{Generalized magnetic mirrors,} Phys. Rev. Lett. \textbf{119},
  123902 (2017).

\bibitem{Bohren1983_book}
C.~F. Bohren and D.~R. Huffman, \emph{Absorption and Scattering of Light by
  Small Particles} (Wiley, 1983).

\bibitem{DOICU_light_2006}
A.~Doicu, T.~Wriedt, and Y.~A. Eremin, \emph{Light Scattering by Systems of
  Particles: Null-Field Method with Discrete Sources: Theory and Programs},
  vol. 124 ({Springer}, 2006).

\bibitem{JACKSON_1998__Classical}
J.~D. Jackson, \emph{Classical {{Electrodynamics Third Edition}}} ({Wiley},
  {New York}, 1998), 3rd ed.

\bibitem{LALANNE__LaserPhotonicsRev._Light}
P.~Lalanne, W.~Yan, K.~Vynck, C.~Sauvan, and J.-P. Hugonin, \enquote{Light
  {{Interaction}} with {{Photonic}} and {{Plasmonic Resonances}},} Laser
  Photonics Rev. \textbf{12}, 1700113 (2018).

\bibitem{KRISTENSEN_2020_Adv.Opt.Photon.AOP_Modeling}
P.~T. Kristensen, K.~Herrmann, F.~Intravaia, and K.~Busch, \enquote{Modeling
  electromagnetic resonators using quasinormal modes,} Adv. Opt. Photon., AOP
  \textbf{12}, 612--708 (2020).

\bibitem{CHEN_Phys.Rev.Lett._Extremize}
W.~Chen, Q.~Yang, Y.~Chen, and W.~Liu, \enquote{Extremize {{Optical
  Chiralities}} through {{Polarization Singularities}},} Phys. Rev. Lett.
  \textbf{126}, 253901 (2021).

\bibitem{WANG_2024_Phys.Rev.Lett._Geometric}
P.~Wang, Y.~Chen, and W.~Liu, \enquote{Geometric {{Phase-Driven Scattering
  Evolutions}},} Phys. Rev. Lett. \textbf{133}, 093801 (2024).

\bibitem{YARIV_2006__Photonics}
A.~Yariv and P.~Yeh, \emph{Photonics: {{Optical Electronics}} in {{Modern
  Communications}}} ({Oxford University Press}, {New York}, 2006), 6th ed.

\bibitem{Johnson1972_PRB}
P.~B. Johnson and R.~W. Christy, \enquote{Optical constants of the noble
  metals,} Phys. Rev. B \textbf{6}, 4370 (1972).

\end{thebibliography}

\end{document}